\documentclass[10pt,conference]{./IEEEtran}
\usepackage{cite}
\usepackage{amsmath,amssymb,amsfonts}
\usepackage{algorithmic}
\usepackage{graphicx}
\usepackage{textcomp}
\usepackage{xcolor}
\def\BibTeX{{\rm B\kern-.05em{\sc i\kern-.025em b}\kern-.08em
		T\kern-.1667em\lower.7ex\hbox{E}\kern-.125emX}}

\usepackage{multirow}
\usepackage{subfig}

\usepackage{pgfplots}
\pgfplotsset{compat=newest}
\usepgfplotslibrary{units}
\usepgfplotslibrary{groupplots}
\pgfplotsset{%
    every axis/.style={%
        axis line shift=2ex,
        axis lines*=left,
        enlargelimits=false,
    },
    set layers,
    every axis/.append style={%
        xlabel={x},
        ylabel={y},
        ylabel shift={-.1cm},
        xmajorgrids=true,
        ymajorgrids=true,
        width=\linewidth,
        height=5cm,
        scaled y ticks=false,
        y tick label style={font=\footnotesize},
        x tick label style={font=\footnotesize},
        log ticks with fixed point,
        legend pos=north east,
        legend cell align=left,
        legend style={draw=none},%, font=\tiny},
        cycle list={%
            {black, mark=*},
            {red, mark=square*},
            {blue, mark=triangle*},
            {cyan, mark=halfsquare*, mark color=cyan},
            {orange, mark=pentagon*}
        },
        bar cycle list/.style={cycle list={%
                {black, fill=black!75, mark=none},
                {red, fill=red!75, mark=none},
                {blue, fill=blue!75, mark=none},
                {cyan, fill=cyan!75, mark=none},
                {orange, fill=orange!75, mark=none}
            },
        },
    },
}

\usepackage[disable]{todonotes}
\IEEEoverridecommandlockouts\IEEEpubid{\makebox[\columnwidth]{978-1-7281-4973-8/20/\$31.00 ~\copyright~2020 IEEE \hfill} \hspace{\columnsep}\makebox[\columnwidth]{ }}

\begin{document}

\title{Scan Correlation -- Revealing distributed scan campaigns}

\author{
	\IEEEauthorblockN{1\textsuperscript{st} Steffen Haas}
	\IEEEauthorblockA{\textit{University Hamburg, Germany} \\
		haas@informatik.uni-hamburg.de}
	\and
	\IEEEauthorblockN{2\textsuperscript{nd} Florian Wilkens}
	\IEEEauthorblockA{\textit{University Hamburg, Germany} \\
		wilkens@informatik.uni-hamburg.de}
	\and
	\IEEEauthorblockN{3\textsuperscript{rd} Mathias Fischer}
	\IEEEauthorblockA{\textit{University Hamburg, Germany} \\
		mfischer@informatik.uni-hamburg.de}
}

\maketitle

\begin{abstract}
Public networks are exposed to port scans from the Internet. Attackers search for vulnerable services they can exploit. In large scan campaigns, attackers often utilize different machines to perform distributed scans, which impedes their detection and might also camouflage the actual goal of the scanning campaign. 
In this paper, we present a correlation algorithm to detect scans, identify potential relations among them, and reassemble them to larger campaigns. We evaluate our approach on real-world Internet traffic and our results indicate that it can summarize and characterize standalone and distributed scan campaigns based on their tools and intention. 
%The results are summaries to characterize standalone and distributed campaigns regarding their techniques and intention.
%For evaluation, we demonstrate findings with our approach on real-world Internet traffic.
\end{abstract}

\begin{IEEEkeywords}
    Intrusion Detection, Alert Correlation, Port Scans, Distributed Attacks
\end{IEEEkeywords}

%%% Beginn des Artikeltexts
\section{Introduction}
% Context
%\begin{itemize}
%    \item Port Scans are an early stage to prepare a following attack
%    \item Networks are constantly scanned from the Internet
%    \item Important to understand what attackers scan for to derive and prepare appropriate countermeasures
%\end{itemize}
Port scans are used by attackers for reconnaissance to prepare attacks and to find a viable entrypoint into a target network or system~\cite{al2014network} or to explore an infiltrated network through lateral movement~\cite{wilkens2019towards}. For that, attackers try to identify and enumerate running Internet services by scanning for open ports on the potential victims. This affects organizations of any size, including Internet Service Providers (ISPs) and cloud providers~\cite{riquet2012large}. Therefore, it is important to detect port scans and to understand the attacker's perspective to predict consecutive attack steps that can then be mitigated at an early stage, especially in the presence of long-running APT attacks~\cite{chen2014study}.

% Problem
%\begin{itemize}
%    \item Volume of scanned ports and systems
%    \item Different intentions and proceeding of attackers
%    \item Coordinated scans where scan tasks are distributed among attackers
%\end{itemize}
Gaining intelligence about these scans is complicated as different attackers usually perform overlapping scans in parallel, whereas some sophisticated attackers even cooperatively scan networks from multiple machines making it harder to link those. To achieve actionable threat intelligence and to be able to react to unusual or new scanning behavior, a methodology is needed that captures characteristics of attackers and to enable the correlation of related scans and distributed scanning nodes.

% Existing Solutions
%%% Skip?

% Contribution
%\begin{itemize}
%    \item Detection of Port Scan Campaigns for standalone and colluding attackers that describes the characteristics of related scans as summary of the campaign
%    \item Identification of features that identify colluding attackers, both regarding knowledge gain and used tools
%\end{itemize}
Our contribution is a methodology to detect port scan campaigns by correlating scans activity from standalone or distributed scanners. Our approach is based on the key insight that scan activity from the same attacker exposes similar properties, even when accomplished by a coordinated scan using multiple nodes. Thus, we describe features that identify related scans in both knowledge obtained by the attacker as well as the choice of target hosts. Our approach uses these features to derive similarity scores between single scans and then applies thresholding to label the resulting clusters as distributed port scan campaigns.

% Results
%\begin{itemize}
%   \item
%\end{itemize}
We evaluate our work on real-world traffic in Internet backbone, which contains a broad range of port scans from multiple attackers. We present our findings on scan campaigns and especially those that are performed by distributed scanners. Furthermore, we show that larger distributed campaigns can already be detected with the visibility of enterprise networks.

% Outline
The remainder of the paper is structured as follows. In Section~\ref{sec:related_work} we describe related work. Section~\ref{sec:system} describes our approach to reveal standalone and distributed scan campaigns. In Section~\ref{sec:evaluation} we evaluate our approach and Section~\ref{sec:conclusion} concludes our paper.

\section{Related Work}
\label{sec:related_work}

% Classification
Apart from reviewing approaches specifically for the detection of port scans, we also review approaches for correlating potentially Internet-wide attack information in large network.

\paragraph{Port Scan Detection}
% Practical automated detection of stealthy portscans
In~\cite{staniford2002practical}, Staniford et al. describe \emph{Spice} an approach to cluster anomalous packets to port scans. Spice is focused on long and stealthy port scans that are not caught by simple detection mechanisms as the single scan probes just fall out of the detection window. To combat this the authors do anomaly detection on the packets first and then keep those with higher scores for longer periods of time. However, although the authors mention the problem of colluding scanners and claim that Spice can detect those, it is not the focus of the approach and partially unclear how they achieve detection.

% Detection and Characterization of Port Scan Attacks -> Port characterization derived but more general definition of colluding attackers
Lee et al.~\cite{lee2003portscan} propose an approach to detect distributed port scans. They classify port scans as either \emph{vertical} (scanning multiple ports of the same host), \emph{horizontal} (scanning the same port on multiple hosts) or \emph{block scans} (the combination of the two). We use a similar metric when characterizing the source and destination ports of attackers. However their definition of colluding scanners is limited to the source IPs being in the same \texttt{/24} subnet to account for decoy or zombie machines. Our approach uses a larger set of ten distinct features to calculate similarity between attacking IPs and is thus not limited to this assumption.

% Coordinated scan detection -> No notion of camouflage and larger feature set
In~\cite{gates2009scan}, Gates also developed an approach to detect distributed port scans. They use a set covering approach based on a feature set of number of targeted ports, number of targeted IPs, the selection algorithm for the target IPs and how additional IPs were scanned to hide the true target, if any. The author chose this set to model the attackers' incentives rather than specifically defining how a distributed scan looks like. Our approach is similar but additionally incorporates indicators for same tools and techniques used in distributed port scans.

% Internet-wide view of Internet-wide scanning ->
Durumeric et al.~\cite{durumeric2014internet} conduct a real-world measurement on Internet-wide scanning. For that, they propose heuristics for recognizing large horizontal scans by leveraging fingerprinting of well-known scanning tools, including \emph{ZMap}~\cite{durumeric2013zmap} and \emph{Masscan}. Their results indicate that the current practice of Internet-wide scanning is indeed mostly horizontal. Furthermore, the authors analyze the scan campaigns semantically regarding their goals.
In contrast to them, we apply a more flexible detection approach that focuses on the detection of distributed scanners that collude in a larger scan campaign.

\paragraph{Network-wide Alert Correlation}

% Similarity thresholds
Feature similarity -- as leveraged by our approach -- has been used multiple times for intrusion detection and more specifically for alert correlation~\cite{valdes2001probabilistic,xie2004anomaly,zhou2009collaborative,debar2015aggreagtion}. 
% GAC: graph-based alert correlation for the detection of distributed multi-step attacks
However, common alert clustering as in \emph{GAC}~\cite{haas2018gac} is too generic to capture the intrinsic characteristics of distributed port scans. Scan-specific features are required for high evidence about colluding scanners or otherwise, the individual scans are not assembled to a scan campaign.

% CIDS
When dealing with potentially Internet-wide scan campaigns, their detection faces the locality problem~\cite{stein2017classification}, i.e., the local vision a specific Intrusion Detection System (IDS) has on the monitored network. To overcome this problem, the sensors in a Collaborative Intrusion Detection System (CIDS)~\cite{vasilomanolakis2015taxonomy} exchange data to establish a holistic view on a larger scope.
% SkipMon: A Locality-Aware Collaborative Intrusion Detection System
In particular, the CIDS \emph{SkipMon}~\cite{vasilomanolakis2015skipmon} identifies monitors with similar traffic to group local monitors. However, traffic similarity is not a sufficient measure to identify similar scan activity that results from distributed scanners.

\section{Detection of Scan Campaigns}
\label{sec:system}

% Intro
Massive port scan activity from the Internet results in a sheer flood of IDS alerts for the scanned hosts and ports. In this section, we present our alert correlation algorithm to detect related scan alerts and summarize them as scan campaigns.
% Overview
For that, the algorithm first characterizes the scan activity regarding its techniques and intention. Afterwards, scanners with similar behavior are eventually classified as coordinated by the same attacker to collectively gain some knowledge about a target network. The result is a description of scan campaigns that summarize the holistic scan activity of an attacker.

\subsection{Port Scan Detection}
\label{subsec:system_scandetetion}

% Network Communication and Connections
%\begin{itemize}
%   \item Definition of term connections and its purpose
%   \item Three-way handshake and discuss technical reasons for failure
%\end{itemize}
For the transport protocol TCP, there is an explicit three-way handshake~\cite{postel1981rfc} to initiate a connection-oriented communication. More generally, with the term \textit{connection} we refer to a series of related packets exchanged between two endpoints, i.e., IP and port tuples.
Ideally, the initiator of a connection reaches out to a service that is known to run behind an open port on the destination host. This eventually leads to a successful connection establishment. However, congestion in the network or downtime of the service can lead to failed establishments without any malicious intention of the initiator.

% Port Scans
%\begin{itemize}
%   \item Knowledge gain about running services
%   \item Scan methods (SYN, Full...)
%\end{itemize}
The intention is different and considered malicious if it is unknown to the initiator whether a service is usually running on a specific IP and port tuple. We denote a single connection attempt to a specific IP:Port to verify the port status as a \textit{scan probe}. A successful connection proves an open port and therefore a running service. However, most of the guessed ports are probably closed, resulting in failed establishments.
Thus, we define a malicious source IP that sends scan probes as a scanning node (\textit{scanner}). The term \textit{port scan} describes all scan probes of a specific scanner.
An IDS usually reports such failed connection attempts as port scan alerts when exceeding a threshold.

% Scan Campaign
%\begin{itemize}
%   \item Intention behind related Scans (Two-dimensions - Hosts and Ports)
%   \item Coordinated scans with colluding attackers
%\end{itemize}
Typically, many ports and hosts are scanned by an attacker in a \textit{scan campaign} to gain a particular knowledge about a target network. To achieve this goal, multiple machines can work together as \textit{distributed scanners}. Their individually gained knowledge is collectively assembled with respect to the intention of the campaign.

\subsection{Characterizing Port Scans}
\label{subsec:system_characterize}
%\begin{itemize}
%   \item Characterization of scan activity per attacker
%   \item Key features
%\end{itemize}

% Two dimensions: Tool and Intention
The scan of a target network can be characterized out of two perspectives: (1) the information that is gained about the target and (2) the techniques used to retrieve the information. This is reflected by the scan intention and the technical tools for the scan campaign, respectively.
% - Tool
In general, we assume attackers to continuously use the same technical tools while scanning and therefore to show a consistent behavior during a campaign.
% - Intention
To perform large campaigns, distributed scanners scan the target network faster or to stay undetected by coordinating their scan activity in a distributed fashion~\cite{dainotti2015analysis}.

% Key features
In the following, we present ten key features that characterize a port scan, i.e., the respective tool used and the attacker's intention. Intuitively, similarity of these features among scans indicate a relation among them. For large sets of scan alerts, these features are evaluated for each scanner. They are summarized in Table~\ref{tab:key_features} and utilized in the second step of our algorithm to correlate port scans.

\begin{table}[t]
    \centering
    \begin{tabular}{r|c|c|c}
        Feature           & Attacker & Target & Values \\
        \hline\hline
        Source Ports      & X        &        & $[\mathcal{S}+p,~\mathcal{F},~\mathcal{M}]$ \\
        \hline
        Destination Ports &          & X      & $[\mathcal{S}+p,~\mathcal{F},~\mathcal{M}]$ \\
        \hline
        Vertical Scan     &          & X      & $[\text{true},~\text{false}+h)]$            \\
        \hline
        Horizontal Scan   &          & X      & $[\text{true},~\text{false}]$               \\
        \hline
        Scan Validation   & X        &        & $[\text{true},~\text{false}]$               \\
        \hline
        IP Version        &          & X      & $[v4,~v6]$                                  \\
        \hline
        Target Hosts      &          & X      & $n \in \mathbf{N}$                          \\
        \hline
        Scans Probes      &          & X      & $n \in \mathbf{N}$                          \\
        \hline
        Source Subnet     & X        &        & $h$                                         \\
        \hline
        Source Location   & X        &        & coordinates $(x, y)$                        \\
    \end{tabular}
    \caption{Key features to characterize scans regarding attacker (i.e. tool) and target (i.e. intention). Values are absolute numbers ($\mathbf{N}$) or in categories of Single ($\mathcal{S}$), Few ($\mathcal{F}$), Multiple ($\mathcal{M}$) with an optional concrete port number $p$ or host IP $h$.\label{tab:key_features}}
\end{table}

\paragraph{Source and Destination Ports}
% Meaning
The involved ports during a scan are of interest because of two reasons. First, the destination port is directly related to the intention of the scan. Second, the usage of the source port across port scans contributes to fingerprinting the scan tool of the attacker.
% Values
Instead of only looking at actual ports $p$, we also assign source and destination ports a category depending on the number of different ports. The label \textit{Single} ($\mathcal{S}$) is for a single port, \textit{Few} ($\mathcal{F}$) for $2$ to $X$ ports, and \textit{Multiple} ($\mathcal{M}$) for more than $X$ ports.
% Distance
The similarity between source ports and destination ports respectively, is $0$ when the categories mismatch. Upon a match, the similarity is $1$ in general. However, for \textit{Single} ports, we also require the port number $p$ to match, otherwise the similarity is $0.5$.

\paragraph{Vertical and Horizontal Scans}
% Meaning
An indirect characteristic of the scan intention is regarding the number of different hosts and the number of different ports.
% Values
If scans encompass more than one target hosts, it is classified as vertical. If scans encompass more than one port on the same target host, it is classified as horizontal.
% Distance
The similarity of $1$ when comparing the vertical and horizontal property respectively, requires a match. The similarity is $0$ otherwise. When the scans target a single hosts, i.e., no vertical scans, we additionally enforce the probes of both scanners to target the exact same host $h$.

\paragraph{Scan Validation}
% Meaning
Not only legitimate connections sometimes break or fail because of technical reasons, also scans eventually fail in detecting a running service although the respective port is open. To compensate false-negative scan results because of packet loss during the scan, scanning tools might try several attempts for the same host and port.
% Values
Thus, we classify scanners to validate their scan results when they perform more than one scan attempt per host and port.
% Distance
Same classifications for two scans lead to a similarity of $1$ and is $0$ otherwise.

\paragraph{IP Version}
% Meaning
% Values
We differentiate the IP protocol versions $4$ and $6$.
% Distance
The version has to match when comparing two attacks for a similarity of $1$. It is $0$ on a mismatch.

\paragraph{Magnitude of Target Hosts and Scans Probes}
% Meaning
Especially in well coordinated scan campaigns, the scan activity is equally distributed among powerful scanner machines.
% Values
When each scanner contributes equally to the knowledge gain, they scan the same amount of hosts and ports, respectively.
% Distance
For statistical and operational reasons, it is unlikely to see the scan activity being perfectly distributed among the scanners. Thus, the similarity of scanned hosts and ports between two port scans is not based on the absolute number $n \in \mathbf{N}$, but on the order of magnitude. For that, the difference between two numbers $a, b$ must be smaller than the smaller number, i.e. $|a-b| < min(a,b)$. In that case, the similarity is $1$ and $0$ otherwise.

\paragraph{Source IP Subnet and Geolocation}
% Meaning
Another indicator for distributed scanners is the proximity between them. This is eventually with respect to organizational or geographical location. Although this does not apply when utilizing a botnet with infected machines around the globe, is likely when using the servers of a specific provider or data center.
% Values
For that, we leverage the source IP of the attacker and the coordinates of its geolocation.
% Distance
Ideally, we would check if two attacker IPs belong to the same subnet, if known. Instead, we generically calculate a linear similarity between $0$ and $1$ depending on the number of equal leading bits in the IP. If two IPs have the same prefix of length 27 bits, their similarity is $0.84$.
For the geolocation, we differentiate between country and coordinates. If coordinates only differ in few degrees, the similarity is $1$. The similarity is $0.5$ if the coordinates are still in the same country and $0$ for different countries.

\subsection{Correlating Port Scans}
\label{subsec:system_correlate}

%\begin{itemize}
%   \item False-Positive filter
%   \item Similarity function
%   \item Clustering to campaigns
%\end{itemize}

% Goal
Based on the characterization of port scans according to the ten key features from Section~\ref{subsec:system_characterize}, we are able to fingerprint scanners and their scans with respect to tools and intention. In the following, we describe the second algorithm part to correlate scanners are coordinated in a scan campaign.

% False-Positive filter
First, we filter false-positive scan alerts that are likely caused by legitimate reasons that happen from time to time. In contrast, malicious port scans cause many connections to fail as they have to probe many hosts and ports to gain the desired knowledge. Thus, we introduce the threshold $\epsilon$ that is the minimum number or scan probes in a port scan. If less, the respective suspicious IP is filtered and not considered for the correlation of port scans.

% Similarity
Two scanners are believed coordinated if the similarity of their fingerprints exceed a threshold $t$. The similarity over the ten key feature is defined as a weighted average:
\begin{equation*}
    \text{sim} = \frac{\sum_{i=1}^{10}{s(i) * w(i)}}{\sum_{i=1}^{10}w(i)}
\end{equation*}
For that, we compare each of the ten features pair-wise for two port scans. The feature similarity $s(i)$ for the features $i \in [1; 10]$ is in the range between $0$ and $1$. To prioritize certain features that indicate distributed scanners with higher certainty, features can be assigned weights $w(i)$ when calculating the similarity among port scans that is $0 \le \text{sim} \le 1$. The threshold $t$ controls if two scans are similar enough regarding their tools and intention to be assumed to belong to the same campaign.

% Clustering
To find distributed scanners in a large scan alert set based on their similarity $\text{sim}$, we leverage hierarchical clustering. This type of clustering benefits from its parameterization as no prediction about number of clusters or their sizes is required. Instead, we incorporate the threshold $t$ that allows to interactively inspect the resulting clusters when varying this parameter.
In particular, we leverage the clustering algorithm \textit{Unweighted Pair Group Method with Arithmetic mean} (UPGMA~\cite{johnson1967hierarchical}) for two reasons. First, the bottom-up property of this clustering algorithm ensures that the most similar scans become clustered first. Second, UPGMA recalculates similarities of merged clusters by averaging the similarities of all contained elements. As most of the key features have only few possible values (cf. Table~\ref{tab:key_features}), this tries to preserve the clusters dominating features when searching new elements to merge. It results in a campaign description that allows to derive an explanation why certain scans have been clustered, i.e. what their most common features are.

\begin{figure*}
	\centering
	\subfloat[Number of ports\label{subfig:ports_number}]{%
		\begin{tikzpicture}
		\def\fileA{data/source_ports.csv}
		\def\fileB{data/destination_ports.csv}
		\begin{axis}[%
		xlabel={Number of unique scanned ports},
		ylabel={Number of scanners},
		only marks,
		legend entries={{Source Ports}, {Destination Ports}},
		ymax=1500,
		xmin=0,
		xmax=50,
		width=.55\linewidth,
		]
		\addplot table {\fileA};
		\addplot table {\fileB};
		\end{axis}
		\end{tikzpicture}
	}
	\subfloat[Classes of ports\label{subfig:ports_classes}]{%
		\def\fileA{data/source_port_classes.csv}
		\def\fileB{data/destination_port_classes.csv}
		\begin{tikzpicture}
		\begin{axis}[%
		xlabel={Classes for $\epsilon = 100$ and $X = 1$},
		ylabel={Number of scanners},
		xtick={0,...,2},
		xticklabels from table={\fileA}{class},
		legend entries={{Src. Ports},{Dst. Ports}},
		width=.4\linewidth,
		%
		%x tick label style={rotate=45, anchor=east},
		ybar,
		ymin=0, ymax=1500,
		enlarge x limits={abs=0.3},
		axis x line*=bottom,    % prevent arrow
		xtick align=inside,
		]
		\addplot table[header=true, x=n, y=attackers] {\fileA};
		\addplot table[header=true, x=n, y=attackers] {\fileB};
		\end{axis}
		\end{tikzpicture}
	}
	\caption{Distribution for number of unique scanned ports among the attackers}\label{fig:ports}
\end{figure*}
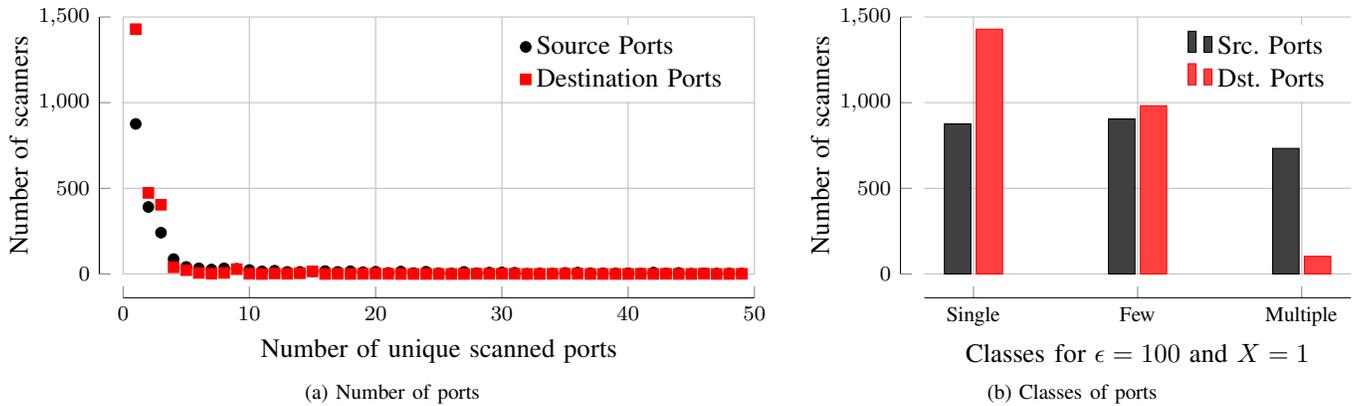

\section{Evaluation}
\label{sec:evaluation}

The evaluation of detecting port scan campaigns is performed on real-world Internet traffic. First, we investigate the data set and the effect of the parameters in our detection. Then, we highlight some of the detected scan campaigns. Furthermore, we investigate the detection accuracy from the perspective of the backbone infrastructure, an ISP, and an enterprise.

\subsection{Data Set}
\label{subsec:eval_dataset}
%\begin{itemize}
%   \item MAWILab: Network Statistics, Zeek for Scan Detection, Scan Alert Statistics
%   \item Restricting Visibility Scope: Backbone, ISP, Company
%\end{itemize}

We first describe the full data set and how we process it to detect port scans. Afterwards, we restrict the network visibility in the data set to reflect the scope for different levels of network providers.

\begin{table}[h]
	\centering
	\begin{tabular}{c|c|c|c}
		\hline
		Scope      & Subnet           & Scan Probes    & Scanners \\
		\hline\hline
		Backbone   & 0.0.0.0/0        & 9,960,652 & 199,403   \\
		ISP        & 133.242.0.0/16   & 2,228,873 & 67,929    \\
		Enterprise & 133.242.179.0/24 & 13,598    & 1,070     \\
	\end{tabular}
	\caption{Port Scan statistics for full data set and simulated network operators\label{tabl:restriction_summary}}
\end{table}

\paragraph{Evaluation Data and Tools}
% Capture
We use a network traffic captured by the MAWILab project~\cite{mawilab}. They monitor a transit cable between the USA and Japan. For anonymity reasons, only IP and TCP/UDP headers are recorded in the network trace. In addition, the last byte of an IP address is randomized consistently among the data set. This potentially effects the key features for the IP and geolocation similarity. However, as the address modification is at the last byte only, the distance between two IPs changes only marginally and the geolocation stays most likely unchanged.

% Scan statistics
We use the network monitor and IDS \textit{Zeek}~\cite{paxson1999bro} to analyze the network capture for TCP port scans, i.e., incomplete TCP establishments. For that, Zeek implements a connection tracking that identifies an incomplete TCP three-way handshake~\cite{postel1981rfc}, e.g., because the destination port is closed or the connection remained half opened. Apart from rare technical reasons like misconfigurations, this usually indicates an attacker that is only interested in discovering open ports, but not in establishing a full connection. In the MAWILab capture of 15 minutes on May, 5th 2019, Zeek detects $9,960,652$ scan probes from $199,403$ scanners to $265,081$ targets.

\paragraph{Restricting Network Visibility}
%  Ground Truth
The MAWILab capture and therefore the detection of the port scans are in the scope of an Internet backbone, as the traffic was recorded from a transit cable. This reflects our ground truth with all the hosts in the data set. Hence, the full data set is the benchmark comparison when it comes to the detection of scan campaigns from the perspective of targeted networks. Different to the holistic view on the scan campaign from the perspective of the Internet backbone, a targeted network might only be one of many targets in the campaign and therefore has only a limited view on the campaign.

To simulate smaller networks, we restrict the visibility of the monitoring and scan correlation to a fraction of IP addresses, i.e., an IP subnet in the full data set. Consequently, the restricted data sets contain only incoming and outgoing communication with respect to the monitored subnet. Apart from the backbone scope, we simulate the scope of an ISP and an enterprise. For the respective subnets, we chose IP addresses that received a lot of scans. The resulting three data sets of different scopes are summarized in Table~\ref{tabl:restriction_summary}.

\subsection{Parameter Study}

\begin{table*}
	\centering
	\begin{tabular}{c||c|c|c||c|c|c}
		\hline
		& \multicolumn{3}{c||}{Campaign from Netherlands} & \multicolumn{3}{c}{Campaign from France} \\
		& Enterprise & ISP       & Backbone & Enterprise & ISP       & Backbone   \\
		\hline\hline
		Scan Probes       & $2449$     & $450,489$ & $1,812,160$ & $20$       & $4,479$   & $18,259$\\
		\hline
		Source Ports      & \multicolumn{3}{c||}{$\mathcal{S}+46960$ and $\mathcal{S}+55776$}  & \multicolumn{3}{c}{$\mathcal{S}+30443$}\\
		Destination Ports & \multicolumn{3}{c||}{$\mathcal{M}$} & \multicolumn{3}{c}{$\mathcal{S}+30443$} \\
		Vertical Scan     & \multicolumn{3}{c||}{$true$} & $true$ and $false$
		& \multicolumn{2}{|c}{$true$} \\
		Horizontal Scan   & \multicolumn{3}{c||}{$true$} & \multicolumn{3}{c}{$false$} \\
		Scan Validation   & \multicolumn{3}{c||}{$false$} & \multicolumn{3}{c}{$false$}\\
		IP version        & \multicolumn{3}{c||}{$IPv4$}  & \multicolumn{3}{c}{$IPv4$}\\
		Target Hosts      & $256$ and $248$
		& $65,205$ and $53,011$
		& $260,299$ and $211,552$ & $1$ to $3$
		& $145$ to $196$
		& $591$ to $799$ \\
		Port Scans        & $1747$ and $702$
		& $343,609$ and $106,880$
		& $1,382,844$ and $429,316$ & $1$ to $3$
		& $145$ to $196$
		& $591$ to $799$\\
		\multirow{2}{*}{Source Subnet}     & \multicolumn{3}{c||}{2} & 15 & 27 & 27\\
		& \multicolumn{3}{c||}{addresses in $185.173.217.208/28$} & \multicolumn{3}{|c}{addresses in $88.138.143.0/27$} \\
		Source Location   & \multicolumn{3}{c}{Netherlands} & \multicolumn{3}{c}{France} \\
	\end{tabular}
	\caption{Summary of two campaigns for different network scopes}\label{tab:campaigns}
\end{table*}

%\begin{itemize}
%   \item Number of different Ports
%   \item False-Positive filter
%   \item Similarity Weights
%   \item Clustering Cutoff
%\end{itemize}

In this section, we discuss the most influencing parameters for the detection of port scan campaigns.
% False-Positive Filter
First, we analyze the threshold $\epsilon$ to filter false-positive scan alerts. Although the average are $50$ probes per scanner, about $90\%$ of the scanners cause $15$ or less probes. Consequently, there exists a few scanners that cause hundreds of probes. For example, for $\epsilon = 5$ only $30\%$ of the scanners are left but they still count for $96\%$ of scan probes. For $\epsilon = 100$ even the remaining $1.26\%$ scanners count for $88\%$ of probes. Generally, this distribution supports our assumption that technical failures occur and results in many false-positive scan alerts that can easily be filtered with a small $\epsilon$.

% Port Classes X
Figure~\ref{fig:ports} plots the distribution of ports among the attackers to decide on the parameter $X$ that is the threshold between the classes $\mathcal{F}$ew and $\mathcal{M}$ultiple.
Figure~\ref{subfig:ports_number} shows how many attackers scan a unique number of ports or use a unique number of ports, respectively. It looks close to a exponential distribution. We set $X = 10$ in further experiments, as for the shape for the distribution function changes around this point.
Figure~\ref{subfig:ports_classes} illustrates the resulting classes for ports when applying the parameters $\epsilon = 100$ and $X = 10$.

\begin{figure}
	\centering
	\begin{tikzpicture}
	\def\fileA{data/threshold.csv}
	\begin{axis}[%
	xlabel={Threshold parameter $t$},
	ylabel={Number of campaigns},
	only marks,
	%sharp plot,
	]
	\addplot table[x=t, y=clusters] {\fileA};
	\end{axis}
	\end{tikzpicture}
	\caption{Resulting clusters depending on the parameter $t$\label{fig:clustering_cutoff}}
\end{figure}
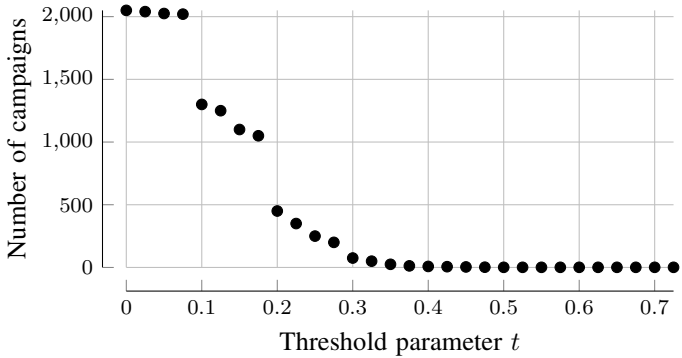

% Similarity Weights
We design the similarity weights $w(i)$ with respect to how strongly similar features $i$ indicate related port scans. We assign the features three different weights as follows. The strongest weight of $4$ is for the two ports. Medium strength with a weight of $2$ is assigned to the vertical and horizontal property, scan verification, and source IP and location. The lowest weight of $1$ is assigned to the IP version, and the order of magnitude for target hosts and scan probes.

% Clustering Cutoff
The effect of the parameter $t$ for the cutoff in hierarchical clustering is illustrated in Figure~\ref{fig:clustering_cutoff}. We set $t = 0.15$ as a reasonable value which is in between two points where the shape significantly changes.

Next, we analyze the resulting clusters, i.e., scan campaigns, for the parameters $X = 10$, $\epsilon$ between $0$ and $10$, and $t = 0.15$.

\subsection{Campaign Characteristics}

% Campaign Statistics
Applying the correlation of port scans (cf. Section~\ref{subsec:system_correlate}) on our data set, results in the campaigns summarized in Table~\ref{tab:campaign_dataset}. As the simulated scope of an ISP and enterprise network restricts the visibility, the number of port scans decreases to $23\%$ and $0.14\%$ respectively compared to the ground truth in the backbone scope. Consequently, also the number of detected campaigns decreases to $10\%$ and $0.76\%$, respectively.

\begin{table}[h]
    \centering
    \begin{tabular}{c|c|r|r|c|rl}
        \hline
        \multirow{2}{*}{Scope}      & Filter & \multirow{2}{*}{Scans}     & \hspace*{-2px}\multirow{2}{*}{Attackers}\hspace*{-2px} & \hspace*{-2px}\multirow{2}{*}{Campaigns}\hspace*{-2px} & \multicolumn{2}{c}{Distributed} \\
        & $\epsilon$ &  &  &  & \multicolumn{2}{c}{Scanners} \\
        \hline\hline
        \hspace*{-2px}Backbone   & 10                & 9,401,543 & 27,244 & 1955 & 19,373 & \hspace*{-10px}(71\%)\hspace*{-2px} \\
        \hspace*{-2px}ISP        & 5                 & 2,130,215 & 8,617  & 717  & 2828   & \hspace*{-10px}(33\%)\hspace*{-2px} \\
        \hspace*{-2px}Enterprise & 0                 & 13,598    & 1070   & 32   & 208    & \hspace*{-10px}(20\%)\hspace*{-2px} \\
    \end{tabular}
    \caption{Summary of campaign detection for clustering parameter $t = 0.15$ for different network scopes\label{tab:campaign_dataset}}
\end{table}

% Detection Accuracy
While the decrease of attackers, scanned ports, and campaigns is directly caused by the restricted visibility scope, it is of interest how this effects the correlation of monitored scans to campaigns. The fraction of scanners that are coordinated, indicates that the detection rate suffers from a restricted monitoring scope. However, this is eventually mostly caused by the nature of restricted visibility, especially when the campaign is not targeting our simulated subnet in particular but it is a broad Internet scan.

% Example campaigns
Out of the 1955 detected campaigns, we look at two specific ones. They represent many other campaigns that we found in the data set. Even though these two are not of more interest than others, we picked them to exemplary highlight the clustering outcome as summary of a scan campaign. In particular, we look at the outcome of the same two campaigns on all three network scopes. The fingerprints of the campaigns are shown in Table~\ref{tab:campaigns}. Apart from the number of scan probes, the table lists the values of the ten key features with respect to Table~\ref{tab:key_features}. This illustrates feature values that are equal for all visibility scopes and others that differ among the scopes.

The two examples show campaigns from Netherlands and from France. The key features show that many values are equal among the distributed scanners. Especially that in the France campaign 27 out of 32 possible IP addresses in the $88.138.143.0/27$ network show the same scanning properties, is a very strong indication for their coordination.
For the Netherlands campaign, we see only two scanners that most probably belong to the same organizational subnet. Although the two different source ports might not be intuitive, from the classification for using a single source port only we derive that the scan tool uses a static port that is not hard-coded but chosen at run-time.
%- Different Scopes
In case of these two example campaigns, we have seen that the correlation of distributed scanners is likely to be successful as long as a sufficiently large number of scan activities from a particular campaign is monitored.

\section{Conclusion}
\label{sec:conclusion}

% System
In this paper, we presented an approach to detect port scan campaigns that potentially deploy distributed scanners. For that, our correlation algorithm for scan alerts detects and fingerprints port scans and identifies those that belong to the same campaign. We presented key features to characterize port scans and to summarize a scan campaign.
% Results
The evaluation on real-world Internet traffic gives insights into the parameterization of our correlation algorithm and into common patterns of port scans. Furthermore, the results highlight the detection of distributed scanners and how the size of the monitored network effects the detection performance.
The results indicate that also small and medium networks can detect scan campaigns on their own when a majority of their hosts is affected.
% Future Work
Future work can identify more features that are either directly available in the packets like the Time To Live (TTL) field or derived like the Autonomous System Number (ASN). Furthermore, the detection of scan campaigns can leverage expert knowledge about fingerprints of well-known scanning tools.

\bibliographystyle{./IEEEtran}
\bibliography{./haas20scans}
\end{document}